\documentclass[twoside]{article}

\usepackage{tabulary,graphicx,times,caption,fancyhdr,amsfonts,amssymb,amsbsy,latexsym,amsmath}
\usepackage[utf8]{inputenc}
\usepackage[nointegrals]{wasysym}
\usepackage{url,multirow,morefloats,cancel,tfrupee,textcomp,colortbl,xcolor,pifont}
\usepackage{amssymb}
\usepackage{gensymb,mathcomp}
\usepackage{slashed}
\usepackage{graphics}
\usepackage{graphicx}
\usepackage{multirow}
\usepackage{epsfig}
\usepackage[tight,hang]{subfigure}

\newcommand{\be}{\begin{equation}}
	\newcommand{\ee}{\end{equation}}
\newcommand{\bea}{\begin{eqnarray}}
	\newcommand{\eea}{\end{eqnarray}}
\newcommand{\bdm}{\begin{displaymath}}
	\newcommand{\edm}{\end{displaymath}}
\newcommand{\baa}{\begin{array}}
	\newcommand{\eaa}{\end{array}}
\newcommand{\ds}{\displaystyle}

\makeatletter

\usepackage{ifxetex}
\ifxetex\else
  \usepackage{dblfloatfix}
\fi

\@ifundefined{subparagraph}{
\def\subparagraph{\@startsection{paragraph}{5}{2\parindent}{0ex plus 0.1ex minus 0.1ex}%
{0ex}{\normalfont\small\itshape}}%
}{}

\def\URL#1#2{\@ifundefined{href}{#2}{\href{#1}{#2}}}

\def\UrlOrds{\do\*\do\-\do\~\do\'\do\"\do\-}%
\g@addto@macro{\UrlBreaks}{\UrlOrds}

\makeatother

\usepackage[paperheight=11in,paperwidth=8.3in,margin=2.5cm,headsep=.7cm,top=2.5cm]{geometry}
\usepackage[T1]{fontenc}

\renewenvironment{abstract}
	{\trivlist\item[]\leftskip0pt\par\vskip4pt\noindent
  	\textbf{\abstractname}\mbox{\null}\\}
	{\par\noindent\endtrivlist}

\def\keywords#1{\par\medskip\par\noindent\textbf{Keywords}: #1\par}

 \date{} \emergencystretch 8pt

\makeatletter
\def\author#1{\gdef\@author{\hskip-\tabcolsep%
	\parbox{\textwidth}{\raggedright\bfseries#1\\[1pc]}}}
\def\address[#1]#2{\g@addto@macro\@author{\\\hskip-\tabcolsep\parbox{\textwidth}{\raggedright%
	\normalsize\normalfont\textsuperscript{#1}#2}}}

\def\correspondence#1{\g@addto@macro\@author{\\\hskip-\tabcolsep\parbox{\textwidth}{\raggedright%
	\vspace*{10pt}\normalsize\normalfont~\\#1~\\[12pt]}}}
\def\email#1{\g@addto@macro\@author{\\\hskip-\tabcolsep\parbox{\textwidth}{\raggedright%
	\normalsize\normalfont Emails: #1}}}

\def\title#1{\gdef\@title{\vspace*{-30pt}%
	\raggedright\textbf{\@journaltitle}~\\%
  \raggedright\bfseries\ifx\@articleType\@empty\vspace*{20pt}\else%
  \vspace*{20pt}\@articleType\vspace*{20pt}\\\fi#1}}
\let\@journaltitle\@empty \def\journaltitle#1{\gdef\@journaltitle{{\normalfont\itshape#1}}}
\let\@articleType\@empty \def\articletype#1{\gdef\@articleType{{\normalfont\itshape#1}}}

\let\@runningHead\@empty \def\RunningHead#1{\gdef\@runningHead{{\normalfont #1}}}

\usepackage{fancyhdr}
\fancypagestyle{headings}{\fancyhf{}
  \fancyhead[R]{\itshape\@runningHead}
  \fancyfoot[C]{\thepage}}
\fancypagestyle{plain}{%
	\fancyhf{}
  \fancyfoot[C]{\thepage}}
\makeatother

\usepackage[%
	numbers,sort&compress%
  ]{natbib}

\journaltitle{Preprint for the Research Article}

\begin{document}

\title{Masses and Thermodynamic Quantities of Quarkonia  in Nonrelativistic Quark Model}

\author{C. Ayd{\i}n}
		
\address[]{Department of Physics, Karadeniz Technical University, 61080, Trabzon, Turkey}

\email{coskun@ktu.edu.tr}%

\RunningHead{Masses and Thermodynamic Quantities of Quarkonia  in Nonrelativistic Quark Model}

\maketitle 

\begin{abstract}
The temperature dependent energy, mass and some of thermodynamic quantities of charmonium and bottomonium  have been calculated by solving the radial Schr{\"{o}}dinger equation with the extended Cornell potential at finite temperature using the Nikiforov-Uvarov method. Obtained results accord with experimental data and theoretical results of the previous studies. We have found the spectrum of quarkonia mass strictly decreases at high values of temperature.
\end{abstract}

\section{Introduction}
Heavier mesons also known as quarkonia are constituents of heavy
quarks such as the charmed (c) and bottom(b). The study of the system of the heavy quarkonium plays a very important role for the Standard Model (SM) and also for the quantifiable test of Quantum Chromodynamics (QCD). It is a topic of interest for the high energy physics \cite{1,2,3,4,5,6}. The quarkonia plays an important role in obtaining information on the restoration of spontaneously broken chiral symmetry in a nuclear medium and understanding quark gluon plasma as a new phase of hadronic matter. It was believed that the quarkonia can help us to extract the nature of quark-antiquark interaction at the hadronic scale and play the same role in probing the QCD as the hydrogen atom plan in the atomic physics.

QCD have played an important role in understanding quarkonium spectroscopy since the discovery of charmonium states.There are several equations for quarkonium corresponding to various assumptions, each derived in different theoretical framework. The initial models describing charmonium spectroscopy, using a QCD motivated
Coulomb plus linear confining potential with color magnetic spin dependent interactions have successfully been used. This approach provides a useful framework for refining our
understanding of QCD and also is a guide toward progress in quarkonium physics. Also, all of the open quantum system approaches to quarkonium dynamics have examined  in the quark-gluon plasma.
The Cornell
potential and its extended forms have been successful in describing most known
mesons. In the nonrelativistic quantum mechanic, quarkonia is a bound system and it is described by two particle Schr{\"{o}}dinger equation. The radial Schr{\"{o}}dinger equation can be solved by
the real part of complex value potential and the solution can be used to understand different phenomena in the study of the atomic physics, nuclear physics and
also in high energy physics, which are not yet understood \cite{7}. The radial Schr{\"{o}}dinger equation has been solved by using different methods \cite{8,9,10,11,12}.

One of the methods used in the analytical solution of the Schr{\"{o}}dinger equation is the
Nikiforov-Uvanov method \cite{13}.
This method has been introduced for the solution of the
Schr{\"{o}}dinger equation to its energy levels by transforming it into hyper-geometric type seconder differential equations. It is based on the determination of the wave function solution in terms of special orthogonal functions for any general second order differential equation. Alberico et. al. \cite{14}, Moscy et. al.\cite{15} Agotiya et. al.\cite{7} Abu-Shady et.al.\cite{16} and Ahmadov et.al.\cite{17,18,19,20} have solved the Schr{\"{o}}dinger equation at finite temperature by using a temperature dependent effective potential. 

The study of thermodynamics quantities for quantum systems in different potential has received special attention in the last decades \cite{21,22,23,24}. For a deep understanding of the thermal effects on various physical systems, the energy behavior of the
Schr{\"{o}}dinger equation for a particle in a gas system with an interaction potential and also
the features of particle with respect to thermodynamic properties has to be investigated. Recently, many authors
have studied the thermodynamic properties of particle systems. There are a large number of
potentials used in Schr{\"{o}}dinger, Klein-Gordon or Dirac equations \cite{25,26} with physical importance.

In this study, we consider the extended Cornell potential and calculate the temperature dependent energy, mass and thermodynamic quantities (free energy $F(\beta)$, mean energy $U(\beta)$, specific heat $C(\beta)$, entropy $\mathcal{S}(\beta)$ and magnetic susceptibility $\chi(\beta)$ in terms of canonical partition function $Z(\beta)$) for quarkonia. In contrast to the previous studies using fixed temperature ($T = 0$)for the mass calculations, we calculated mass at varying temperatures. 

The rest of the paper is structured as follows: In Section 2, we present the calculations of the temperature dependent energy eigenvalue expansion and quarkonium mass, calculations of the thermodynamic quantities are given in Section 3. The numerical results and our discussion is provided in Section 4. Finally Section 5. is reserved for the conclusion. 

\section{Temperature Dependent Radial Schr{\"{o}}dinger Equation and Quarkonia Masses}

The radial Schr{\"{o}}dinger equation for two particles ($c=1$ and  $\slashed{h}=1$) interacting via a potential in the three dimensional space is given in \cite{19} as: 
\be \label{teq1} \chi''(r) + 2\mu \left[ E - {1 \over 2\mu} {L (L+1) \over r^2} - V(r) \right] \chi(r) = 0 \ee
where $\mu$ and $L$ are the reduce mass for the quarkonium particle and the angular momentum quantum number, respectively.

We consider the extended Cornell potential, which is the sum of the Cornell potential, the inverse quadratic potential and the harmonic potential.

Our choice for the radial scalar power potential is
\be V(r)=A r-\frac{B}{r}+\frac{C}{r^{2}}+D r^{2} . \ee

Based on the success of the potential of the model at zero temperature, and on the idea that color screening implies modification of the inter-quark forces, potential models have been used to try to understand quarkonia properties at finite temperature \cite{15}. In a thermodynamics environment of interacting quarks and gluons, at temperature $T$,
quark binding becomes modified by color screening. Following the same philosophy for
Cornell potential in \cite{7,27,28,29,30}  and making the potential term as temperature dependent as follows

\be V(r,T)=\mathcal{A}(r,T) r-\frac{\mathcal{B}(r,T)}{r}+\frac{\mathcal{C}(r,T)}{r^{2}}+\mathcal{D}(r,T) r^{2} \ee
where 
\bdm \mathcal{A}(r,T)=\frac{A}{m_{D}(T) r}\left(1-e^{-m_{D}(T) r}\right) \quad, \quad \mathcal{B}(r,T)=B e^{-m_{D}(T) r} \edm
\bdm \mathcal{C}(r,T)=C e^{-m_{D}(T) r} \quad, \quad \mathcal{D}(r,T)=\frac{D}{m_{D}(T) r}\left(1-e^{-m_{D}(T) r}\right) \edm
such that $m_{D}(T)=\gamma \alpha_s(T) T$ is the Debye screening mass which vanishes as $T=0$ and $\alpha_s(T)$ is the running coupling constant 
$\alpha_s(T)=\frac{2\pi}{(11-{2 \over 3} N_f)\ln(T/\Lambda)}$ \cite{31,32,33,34,35,36}. Using the approximation $e^{-m_{D}(T) r} \approx 1-m_{D}(T) r+\frac{1}{2} m_{D}^{2}(T) r^{2}$ up to second order which give a good accuracy for $m_{D}(T) r \ll 1$. Then we obtain
\be \baa{rcl}
V(r,T) & = &\ds B m_{D}(T)+\frac{1}{2} C m_{D}^{2}(T)+\left(A-\frac{1}{2} B m_{D}^{2}(T)\right) r-\left(B+C m_{D}^{2}(T)\right) \frac{1}{r} \\
&+&\ds \left(D-\frac{1}{2} A m_{D}(T)\right) r^{2}+\frac{C}{r^{2}}-\frac{1}{2} D m_{D}(T) r^{3} \\
&=&\ds F+G r-\frac{K}{r}+M r^{2}+\frac{C}{r^{2}}-N r^{3} .

\eaa \ee
If we substitute the value of $V(r,T)$ in Eq(1) we get
\be \label{eqvr} \chi^{\prime \prime}(r)+2 \mu \left[E-\frac{\hbar^{2}}{2} \frac{L(L+1)}{\mu r^{2}}-F-G r+\frac{K}{r}-M r^{2}-\frac{C}{r^{2}}+N r^{3}\right] \chi(r)=0 . \ee
Using $\ds r=\frac{1}{x}$ Eq.(5) becomes:
\be \label{eqnvx}
\hspace{-0.5cm}\chi^{\prime \prime}(x)+\frac{2 x}{x^{2}} \chi^{\prime}(x)+2 \mu \frac{1}{x^{4}}\left[E-\frac{\hbar^{2}}{2 \mu} L(L+1) x^{2}-F-\frac{G}{x}+K x-\frac{M}{x^{2}}-C x^{2}+\frac{N}{x^{3}}\right] \chi(x)=0 .
\ee

Let us consider a characteristic radius of quark and antiquark system, which is the minimum
interval between two quark at which they cannot collide with each other. Setting $y=x-\delta$, we expand the terms $\frac{G}{x}, \frac{M}{x^{2}}$ and $\frac{N}{x^{3}}$ in powers series form around $r_{0}\left(\delta=\frac{1}{r_{0}}\right)$ using up to the second-order terms only, we obtain:
\be \label{eqnG} \frac{G}{x}=G\left(\frac{3}{\delta}-\frac{3 x}{\delta^{2}}+\frac{x}{\delta^{3}}\right) \ee
\be \label{eqnM} \frac{M}{x^{2}}=M\left(\frac{6}{\delta^{2}}-\frac{8 x}{\delta^{3}}+\frac{3 x^{2}}{\delta^{4}}\right) \ee
\be \label{eqnN} \frac{N}{x^{3}}=N\left(\frac{10}{\delta^{3}}-\frac{15 x}{\delta^{4}}+\frac{6 x^{2}}{\delta^{5}}\right) . \ee
This approximation is similar to Pekeris approximation \cite{37,38} that causes deformation of the centrifugal potential.

By substituting Eqs.(7-9) into Eq.(6):
\be \baa{rcl}\ds
\hspace{-0.5cm}\chi^{\prime \prime}(x)+\frac{2 x}{x^{2}} \chi^{\prime}(x) &+& \ds 2 \mu \frac{1}{x^{4}}\left[\left(E-F-\frac{3 G}{\delta}+\frac{3 M}{\delta^{2}}+\frac{10 N}{\delta^{3}}\right)+\left(K-\frac{3 G}{\delta^{2}}-\frac{8 M}{\delta^{3}}-\frac{15 N}{\delta^{4}}\right) x\right.\\
&+&\ds \left.\left(-\frac{1}{2 \mu} L(L+1)-C-\frac{G}{\delta^{3}}+\frac{3 M}{\delta^{4}}+\frac{6 N}{\delta^{5}}\right) x^{2}\right] \chi(x)=0 .
\eaa
\ee

Now, we have the second-order differential equation
\be \ds
\chi^{\prime \prime}(x)+\frac{\widetilde{\tau}(x)}{\sigma(x)} \chi^{\prime}(x)+\frac{\widetilde{\sigma}(x)}{\sigma^{2}(x)} \chi(x)=0
\ee
where $\widetilde{\tau}(x)=2 x, \sigma(x)=x^{2}$ and $\tilde{\sigma}(x)=2 \mu\left(-D_{1}+D_{2} x-D_{3} x^{2}\right)$ such that $\sigma(x)$ and $\tilde{\sigma}(x)$
are polynomials of degree not greater than two and $\widetilde{\tau}(x)$ is a polynomial of degree not greater than one.

Then, we can use the Nikiforov-Uvarov (NU) method for the solution of the above equation. Following similar approaches in in \cite{18,19,20} we obtain the temperature dependent energy eigenvalue expansion as:
\be \label{eqneng} \baa{rcl} \ds
E_{n l} &=& \ds a_{1}+a_{2} m_{D}(T)+a_{3} m_{D}^{2}(T) \\ \ds
&-&\ds 2 \mu \left[\frac{b_{1}+b_{2} m_{D}(T)+b_{3} m_{D}^{2}(T)}{2 n+1+\sqrt{1+4 L(L+1)+c_{1}+c_{2} m_{D}(T)+c_{3} m_{D}^{2}(T)}}\right]^{2}
\eaa \ee
where
\bdm 
a_{1}=\frac{3 A}{\delta}+\frac{6 D}{\delta^{2}} \quad, \quad a_{2}=B-\frac{3 A}{\delta^{2}}-\frac{5 D}{\delta^{3}} \quad, \quad a_{3}=\frac{1}{2} C-\frac{3 B}{2 \delta^{2}}
\edm
\bdm
b_{1}=B+\frac{3 A}{\delta^{2}}+\frac{8 D}{\delta^{3}} \quad, \quad b_{2}=C-\frac{4 A}{\delta^{3}}-\frac{15 D}{2 \delta^{4}} \quad, \quad b_{3}=-\frac{3 B}{2 \delta^{2}} \edm
\bdm
c_{1}=8 \mu\left(C+\frac{A}{\delta^{3}}+\frac{3 D}{\delta^{4}}\right) \quad, \quad c_{2}=-8 \mu\left(\frac{3 A}{\delta^{4}}-\frac{3 D}{\delta^{5}}\right) \quad, \quad c_{3}=-8 \mu\frac{B}{2 \delta^{4}} .
\edm
We would like to note that the results of Eqn. (12) are same as given for $N=3$ in \cite{16} and for $T=0$  in \cite{16,20}. 

In general, the quark and anti-quark bound states are represented by $_n^{2s+1}L_j$ identified with the $J^{PC}$ values with $\overrightarrow{J} = \overrightarrow{L}+ \overrightarrow{S},
\overrightarrow{S}=\overrightarrow{S_q} + \overrightarrow{S_{\overline{q}}}$ parity  $P=(-1)^{L+1}$ and the charge conjugation $C=(-1)^{L+S}$ with $(n,L)$ being the radial
quantum numbers. So, the S-wave ($L=0$) bounds states are represented by $J^{PC}=0^{-+}$ and $1^{--}$, respectively.
The $P$-wave $(L=1)$ states are represented by $J^{PC}=1^{+-}$ with $L=1$ and $S=0$ while $J^{PC}=0^{++}, 1^{++}$ and $2^{++}$ correspond to $L=1$ and $S=1$, respectively.

We derive the mass spectra of the heavy quarkonium systems such as charmonium and bottomonium that have quark and anti-quark of same flavor. The mass of the bound state is much larger than the mass of the quark. The speed of the charm and the bottom quarks in their respective quarkonia is sufficiently small for relativistic effects in these states. For determining the mass
spectra we use the following relation \cite{39,40}
\be \label{eqnmc} M= 2m_q + E_{nl}\ee
where $m_{q}$ is the bare quark mass for quarkonium. Similarly, the  mass spectra of charmonium, $M_c$, and the mass spectra of bottomonium, $M_b$, can be obtained by using  $m_c$ and $m_b$ instead of $m_q$ in Eq.(13), respectively.

\section{Thermodynamic Quantities}
In order to obtain thermodynamic quantities of the non-relativistic particle system, we
should concentrate on the calculation of the canonical partition function $Z(\beta)$ since all
thermodynamic quantities can be obtained from the partition function. The function $Z(\beta)$
at finite temperature $T$ is obtained through the Boltzmann factor as
\be \label{eqnz1} \ds Z(\beta)=\sum_{n=0}^{N} e^{\ds -\beta E_{n l}} \ee

where $\ds \beta=\frac{1}{k_{\beta} T}$ and $k_{\beta}$ is the Boltzmann constant. Substituting Eq.(12) into Eq.(14), we obtain
$$
\ds Z(\beta)=\sum_{n=0}^{N} e^{-\beta\left[C_{1}+\frac{C_{2}}{\left(2 n+C_{3}\right)^{2}}\right]}
$$
where
\bdm
\ds C_1=a_{1}+a_{2} m_{D}(T)+a_{3} m_{D}^{2}(T) \, ,
\edm
\bdm
\ds C_2= 2 \mu \left[b_{1}+b_{2} m_{D}(T)+b_{3} m_{D}^{2}(T)\right]^2
\edm
and
\bdm
\ds C_3=1+\sqrt{1+4 L(L+1)+c_{1}+c_{2} m_{D}(T)+c_{3} m_{D}^{2}(T)} \quad .
\edm
Using the Poisson summation formula \cite{41,42,43,44} with lowest order approximation 
we can obtain the partition function as
\be \ds
Z(\beta)=\frac{1}{2} e^{-\beta C_{1}}\left[e^{\beta \frac{C_{2}}{\left(C_{3}\right)^{2}}}+e^{\beta \frac{C_{2}}{\left(2 \lambda+C_{3}\right)^{2}}}\right]+\ds \int_{0}^{\lambda} e^{-\beta E_{n l}} d n .
\ee
After obtaining the partition function, we can derive the other thermodynamic quantities
free energy $F(\beta)$, mean energy $U(\beta)$, specific heat $C(\beta)$, entropy $\mathcal{S}(\beta)$ and magnetic susceptibility $\chi(\beta)$ in terms of $Z(\beta)$ for $k_{\beta}=1$
\cite{21,24} as:
\be F=-\frac{1}{\beta} \ln (Z) \ee
\be U=-\frac{\partial \ln (Z)}{\partial \beta} \ee
\be C=-\beta^{2} \frac{\partial U}{\partial \beta} \ee
\be \mathcal{S}= \beta^{2} \frac{\partial F}{\partial \beta} . \ee
\be \chi=- \frac{\partial^2 F}{\partial \beta^2} . \ee

\section{Numerical Results and Discussion}

In this section, we provide some numerical results and graphs of the defined functions
using the proposed formulation. Numerical values for the charmonium and bottomonium energy and mass spectra are presented for $\beta=15 \mathrm{GeV}^{-1}$ and $\lambda=10$. We compare our results with the experimentally well-established resonances and using these comparisons we present some predictions for the states which have not been confirmed yet. For the charmonium states, besides the $J/\psi(1s)$ and $\psi(2s)$ resonance, as in suggested in \cite{20}, we predict $3s \rightarrow \psi(4.040)$, 
$4s \rightarrow \psi(4.260)$, $5s \rightarrow \psi(4.410)$ and $6s \rightarrow \psi(4.510)$. Similarly, for the bottomonium, we predict $\Upsilon(1s), \Upsilon(2s), \Upsilon(3s)$ and $\Upsilon(4s)$ resonances.

\begin{table}[!h]
	\caption{Numerical values mass of spectra for charmonium (in $GeV$) for different values of $l$ and $n$ for
		$A=0.196 \, GeV^2$, $B=1.200$, $C=2.95 \times 10^{-3} GeV^{-1}$, $D=1.4 \times 10^{-5} GeV^3$, $\delta=0.232 \, GeV$, $m_{c}=1.209 \, GeV$.}
	\centering 
	{\begin{tabular}{|c|c|c|c|c|c|c|c|c|c|} \hline 
			\multirow{2}{*}{$\mathbf{n}$} & \multicolumn{3}{|c|} {$S$} & \multicolumn{3}{c|} {$P$} & \multicolumn{3}{c|} {$D$} \\
			\cline { 2 - 10 } & Present & Ref. \cite{20} & Exp. \cite{45} & Present & Ref. \cite{20} & Exp. \cite{45} & Present & Ref. \cite{20} & Exp. \cite{45} \\
			\hline
			1 & $3.108$ & $3.098$ & $3.097$ & $3.273$ & $3.256$ & $3.511$ & $3.524$ & $3.505$ & $3.774$ \\
			2 & $3.679$ & $3.687$ & $3.686$ & $3.772$ & $3.780$ & $3.922$ & $3.919$ & $3.922$ & \\
			3 & $4.013$ & $4.042$& $4.039$ & $4.071$ & $4.100$ &  & $4.164$ &  & \\
			4 & $4.225$ &$4.272$ & $4.459$ & $4.263$ & $4.310$ & & $4.326$ & & \\
			5 & $4.367$ &$4.429$ & $4.421$  & $4.394$ & $4.456$ & & $4.438$ & & \\
			6 & $4.468$ & $4.540$ & & $4.792$ & $4.488$ & & $4.520$ & & \\  \hline
		\end{tabular} \label{tbl1}}
\end{table}

In Table  1, we calculated the charmonium mass from $1S$ to $6D$. We have seen that the obtained results depend on the values of the parameters ($A, B, C, D$ and $\delta$) and the results accord with the previous studies \cite{20} and experiment values \cite{45}. The slight differences are due to the effect of the temperature dependent energy eigenvalue expansion in our calculations.

\begin{figure}[!h]
	\centering 
	\includegraphics[width=8cm]{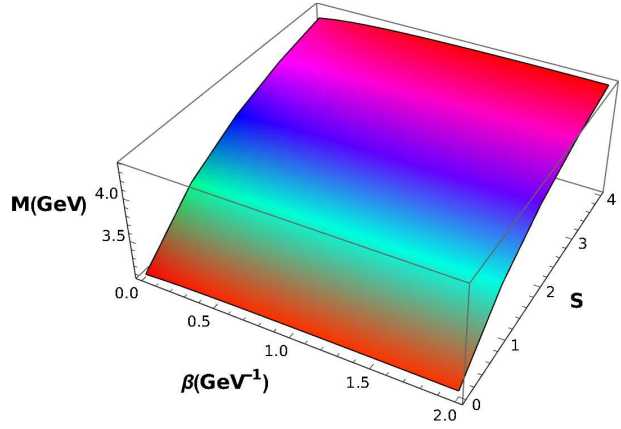}
	\caption{Mass of spectra for the charmonium for $\beta$ change with $S$ values}
	\label{fig1}
\end{figure}

In Figure 1, we presented the mass change of charmonium for changing $\beta$ and $S$ values. It is seen that the increment in $S$ is more effective compared to the increment in $\beta$ in the sense of mass values. 

\begin{figure}[!h]
	\centering 
	\subfigure[$n=1$]{\includegraphics[width=5.3cm]{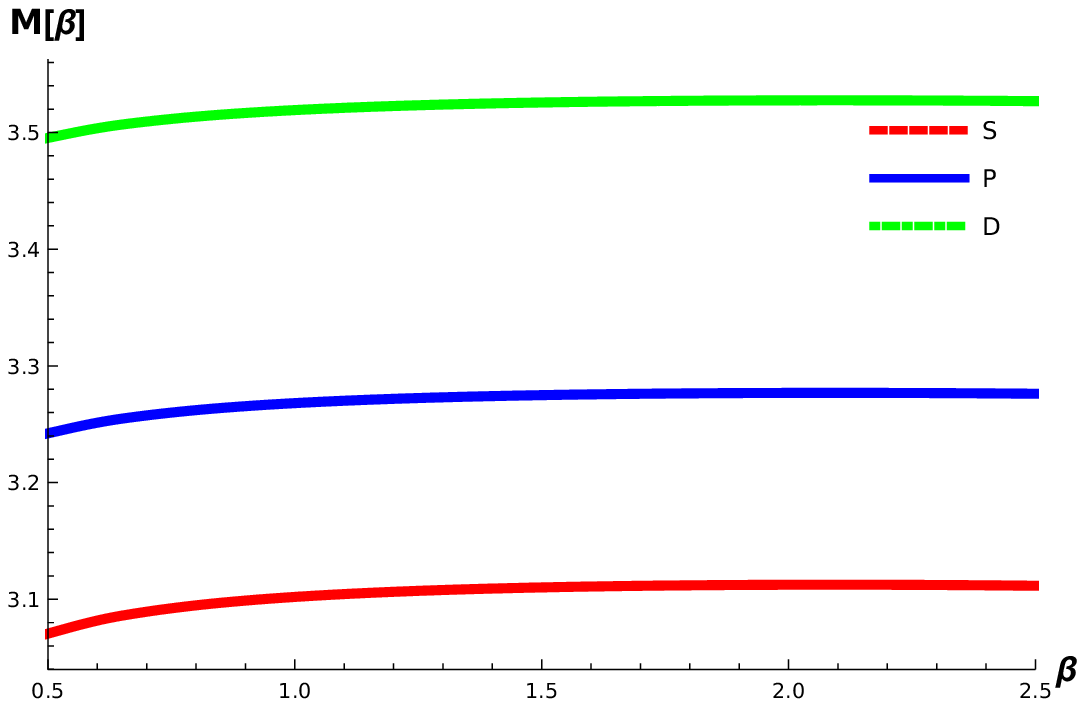}}
	\subfigure[$n=2$]{\includegraphics[width=5.3cm]{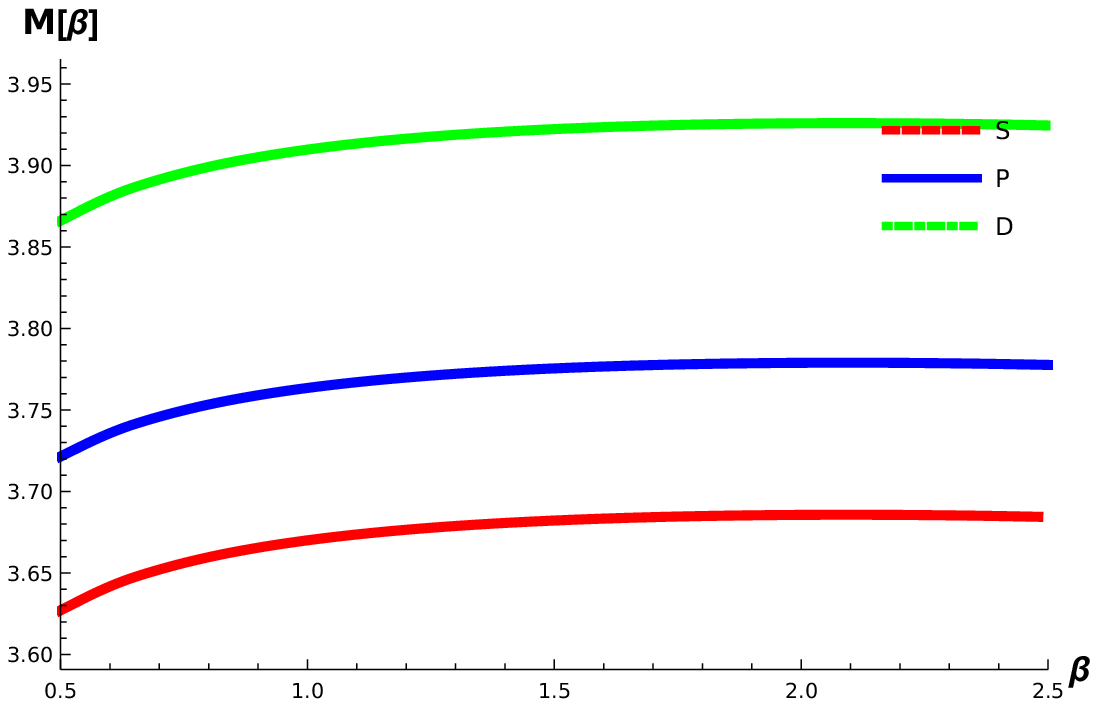}}
	\subfigure[$n=3$]{\includegraphics[width=5.3cm]{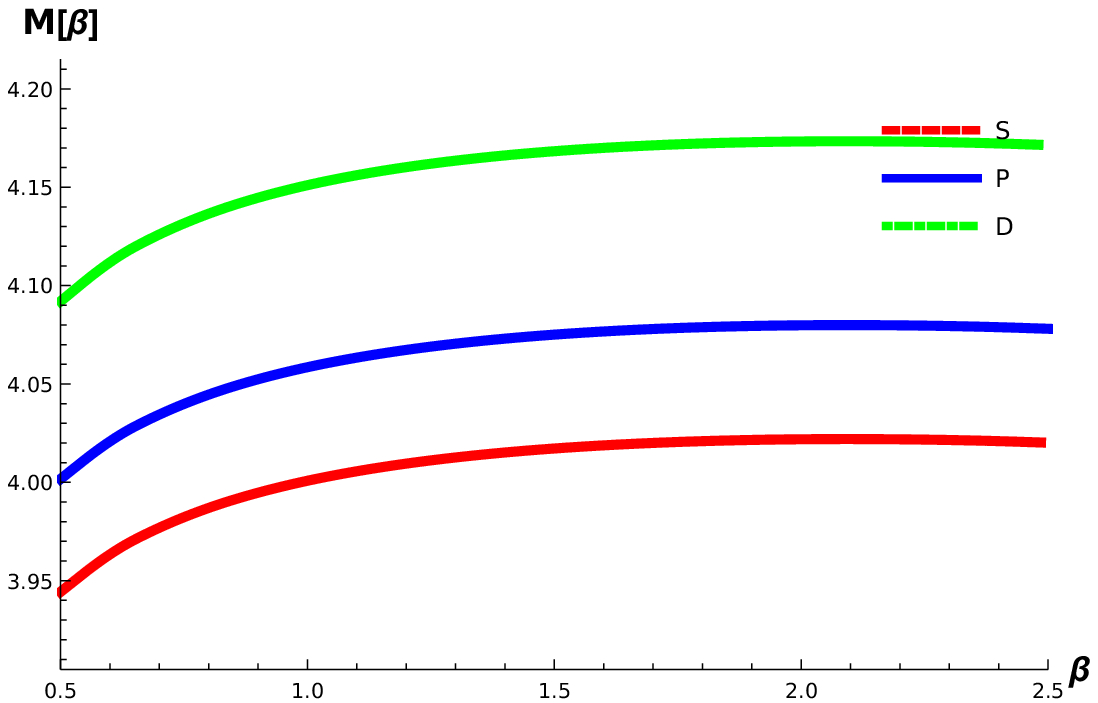}}
	\caption{Mass of spectra for the charmonium for different values of $L$}
	\label{fig2}
\end{figure}

In Figure 2, we compared the mass values for different values of $n$ for $S, P$ and $D$. The value of mass is increasing with increasing $n$ and $L$ values. Also we conclude that the mass converges to its steady value with a sufficiently large $\beta$. Similar behaviors can be observed for the other thermodynamic quantities in Figure 3.

\begin{figure}[!h]
	\centering 
	\subfigure[Partial function]{\includegraphics[width=5.3cm]{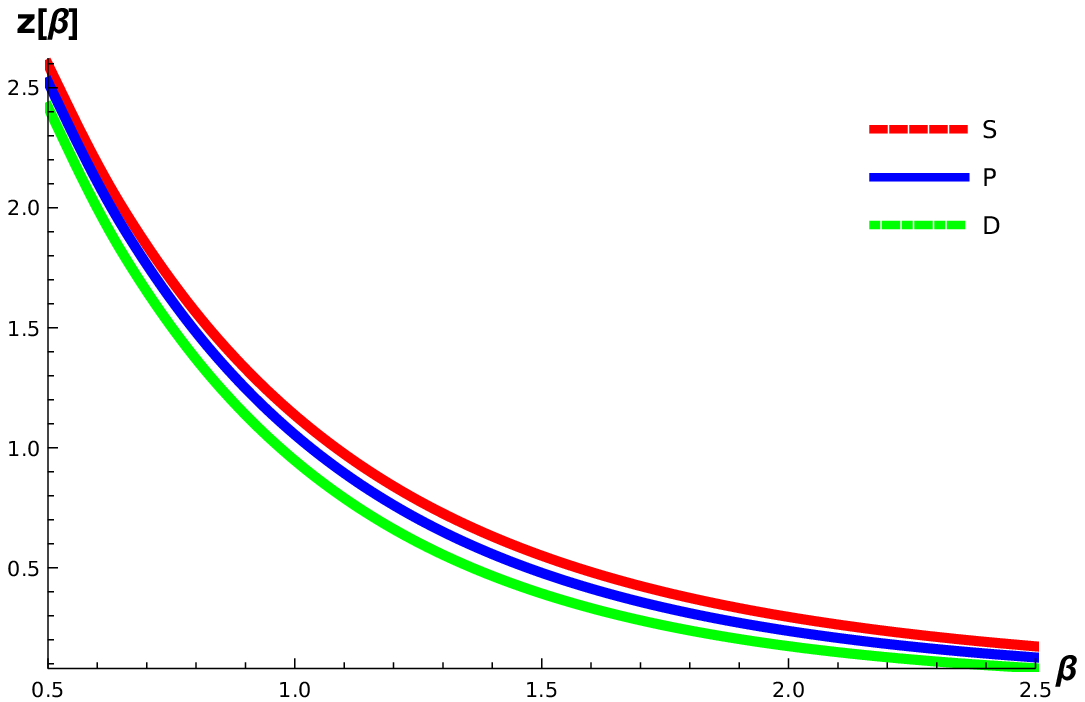}}
	\subfigure[Free Energy]{\includegraphics[width=5.3cm]{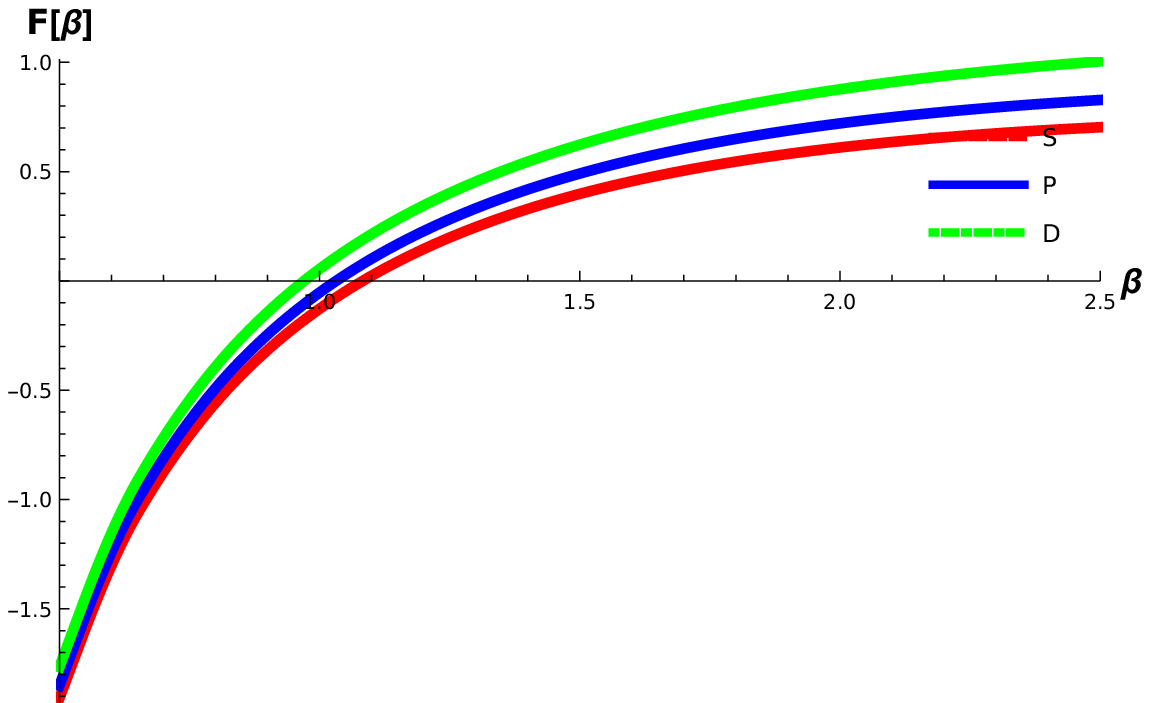}}
	\subfigure[Mean Energy]{\includegraphics[width=5.3cm]{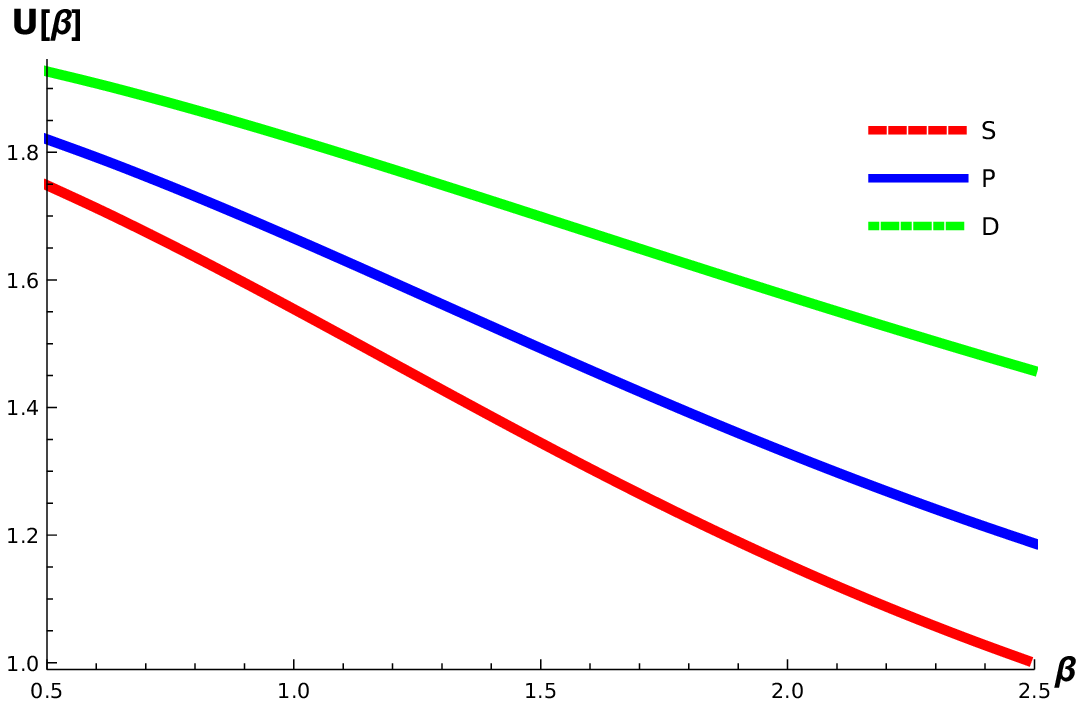}}\\
	\subfigure[Specific Heat]{\includegraphics[width=5.3cm]{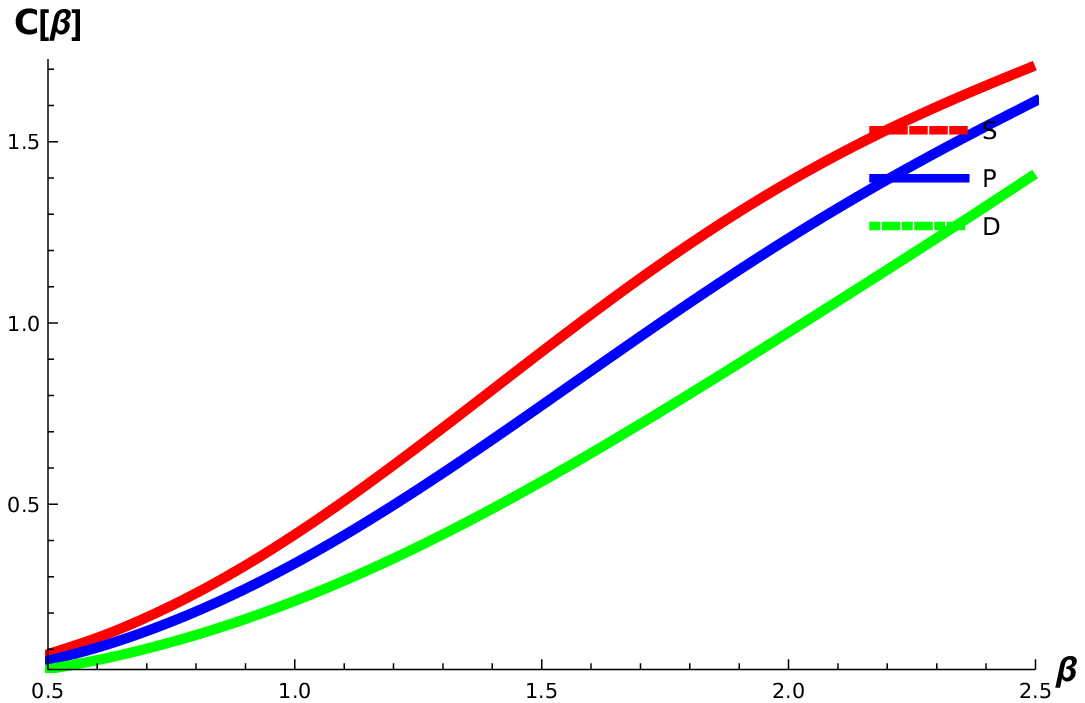}}
	\subfigure[Entropy]{\includegraphics[width=5.3cm]{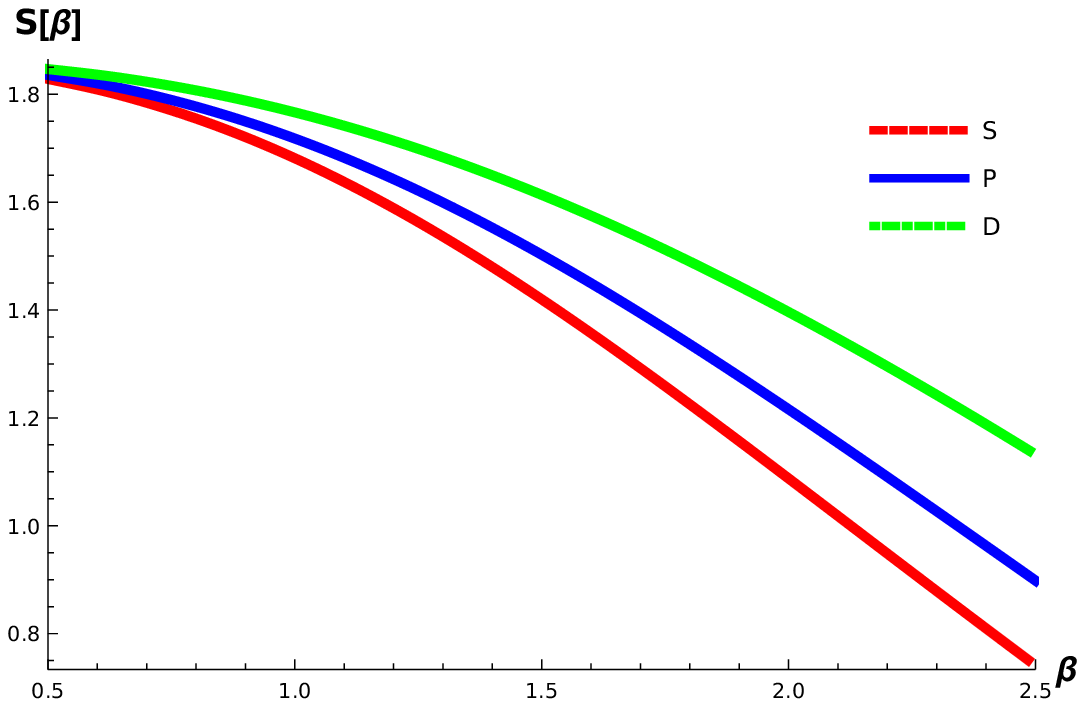}}
	\subfigure[Magnetic Susceptibility]{\includegraphics[width=5.3cm]{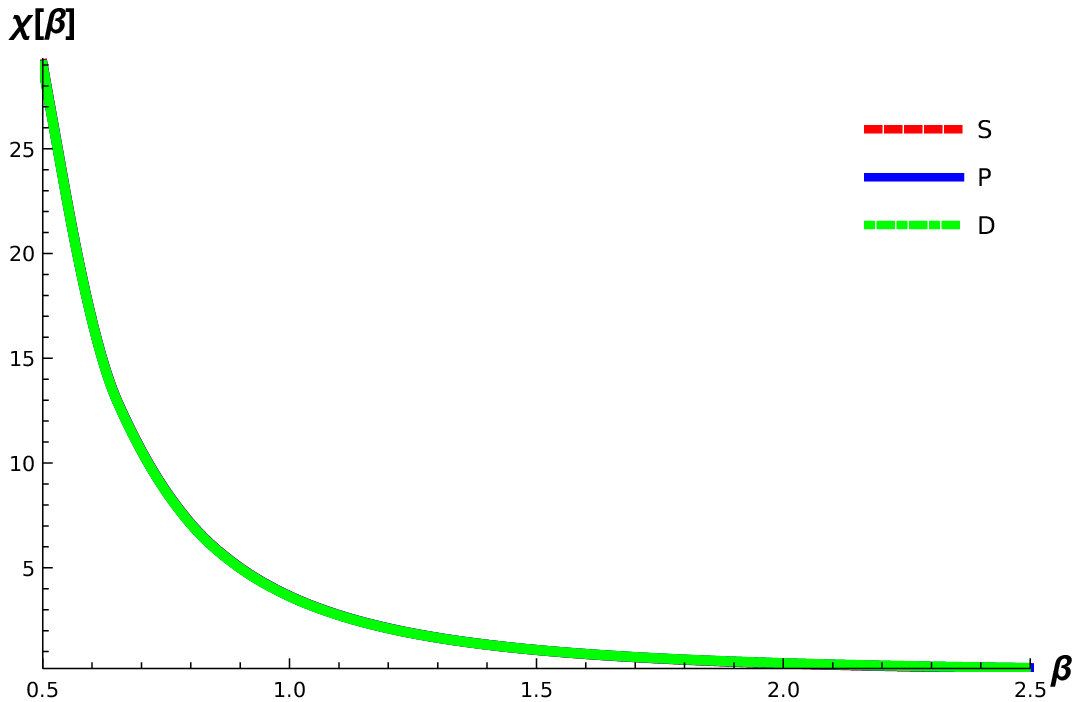}}\\
	\caption{ Thermodynamic quantities for the charmonium for different values of $L$}
	\label{fig3}
\end{figure}

Similar to charmonium, we calculated the mass values for different $L$ and $n$ for botomonium in Table 2. It is seen that the order of the mass values for charmonium and botomonium are almost equal to the order of the bare quark mass of $c$ quark ($m_c$) and the bare quark mass of $b$ quark ($m_b$) and agree with literature results. Due to the effect of the temperature dependent energy eigenvalue expansion formulations, small differences from the reference values are be observed. Also, as stated in \cite{46} by Veliev et. al. the spectrum of bottomonium mass don't change up to $T\simeq100 GeV$ but they start to diminish with increasing the temperature after this point in the framework of the QCD Sum rules.

\begin{table}[!h]
	\caption{Numerical values mass of spectra for bottomonium (in $GeV$) for different values of $l$ and $n$ for
		$A=0.21  \, GeV^2$, $B=1.560$, $C=2.15 \times 10^{-3} GeV^{-1}$, $D=1.4 \times 10^{-5} GeV^3$, $\delta=0.379 \, GeV$, $m_{b}=4.823 \, GeV$.}
	\centering 
	{\begin{tabular}{|c|c|c|c|c|c|c|c|c|c|} \hline
			\multirow{2}{*}{$\mathbf{n}$} & \multicolumn{3}{|c|} {$S$} & \multicolumn{3}{c|} {$P$} & \multicolumn{3}{c|} {$D$} \\
			\cline { 2 - 10 } & Present & Ref. \cite{20} & Exp. \cite{45} & Present & Ref. \cite{20} & Exp. \cite{45} & Present & Ref. \cite{20} & Exp. \cite{45} \\
			\hline
			1 & $ 9.489$ & $9.460$ & $9.460$ & $9.654$ & $9.619$ & $9.899$ & $9.904$ & $9.863$ & $10.164$ \\
			2 & $10.055$ & $10.023$ & $10.023$ & $10.149$ & $10.144$ & $10.294$ & $10.274$ & $$ & $$ \\
			3 & $10.386$ & $10.355$ & $10.355$ & $10.443$ & $10.411$ & $ $ & $10.535$ & $ $ & $ $ \\
			4 & $10.595$ & $10.567$ & $10.579$ & $10.633$ & $10.604$ & $ $ & $10.695$ & $ $ & $ $ \\
			5 & $10.735$ & $10.710$ & $10.876$ & $10.762$ & $$ & $$ & $10.805$ & $$ & $$ \\
			6 & $10.834$ & $$ & $11.019$ & $10.853$ & $$ & $$ & $10.885$ & $$ & $ $ \\ \hline
		\end{tabular} \label{tbl2}}
\end{table}

\begin{figure}[!h]
	\centering 
	\includegraphics[width=8cm]{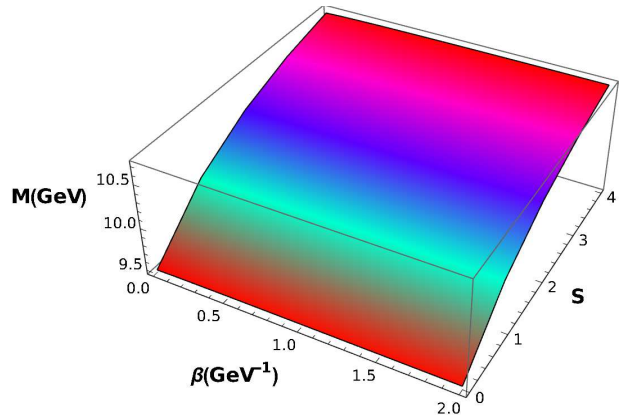}
	\caption{Mass of spectra for the bottomonium for $\beta$ change with $S$ values}
	\label{fig4}
\end{figure}

In Figure 4, we displayed the mass change of bottomonium for changing $\beta$ and $S$ values. The similar behavior as in charmonium is also observed for the bottomonium.

\begin{figure}[!h]
	\centering 
	\subfigure[$n=1$]{\includegraphics[width=5.3cm]{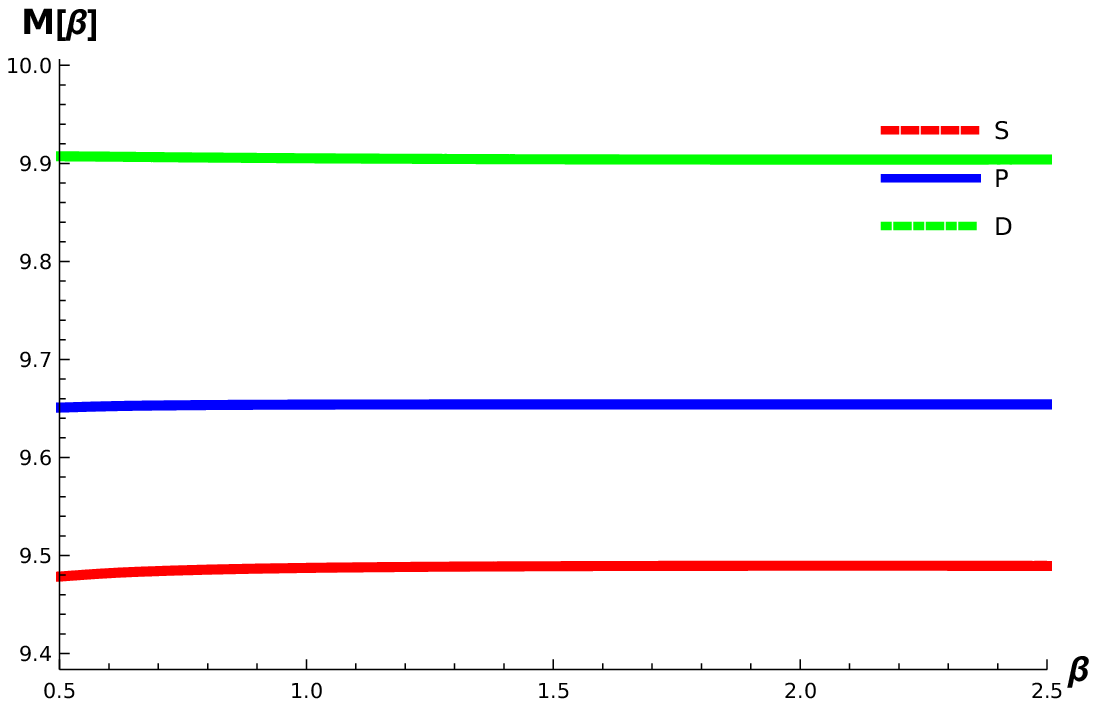}}
	\subfigure[$n=2$]{\includegraphics[width=5.3cm]{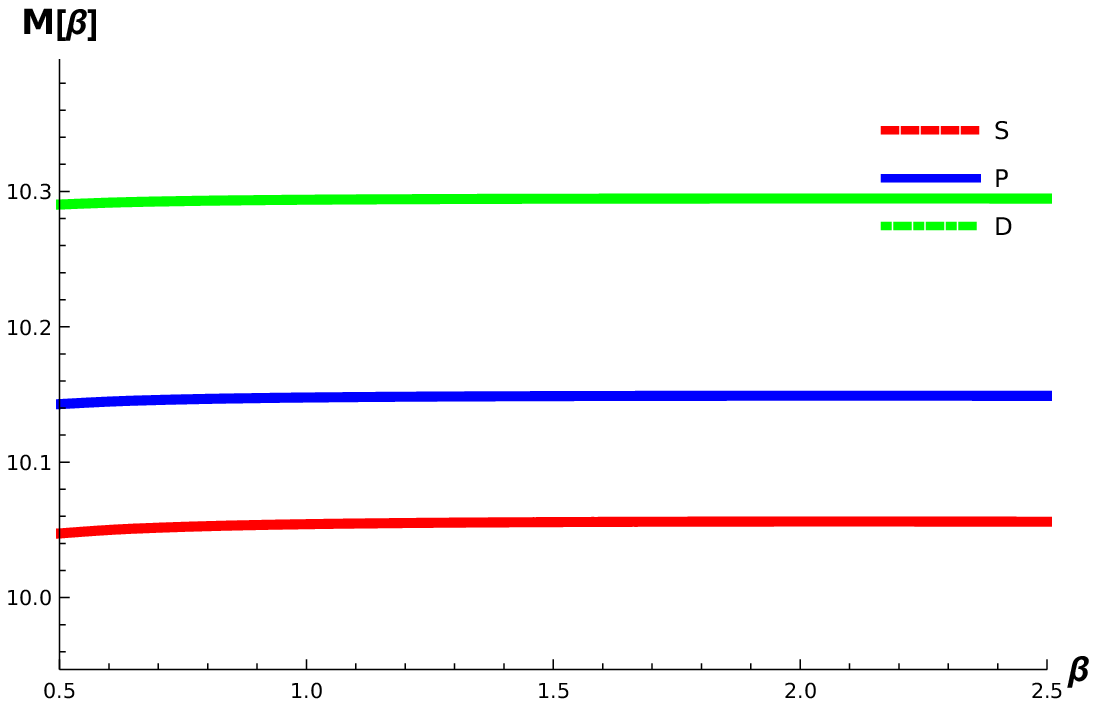}}
	\subfigure[$n=3$]{\includegraphics[width=5.3cm]{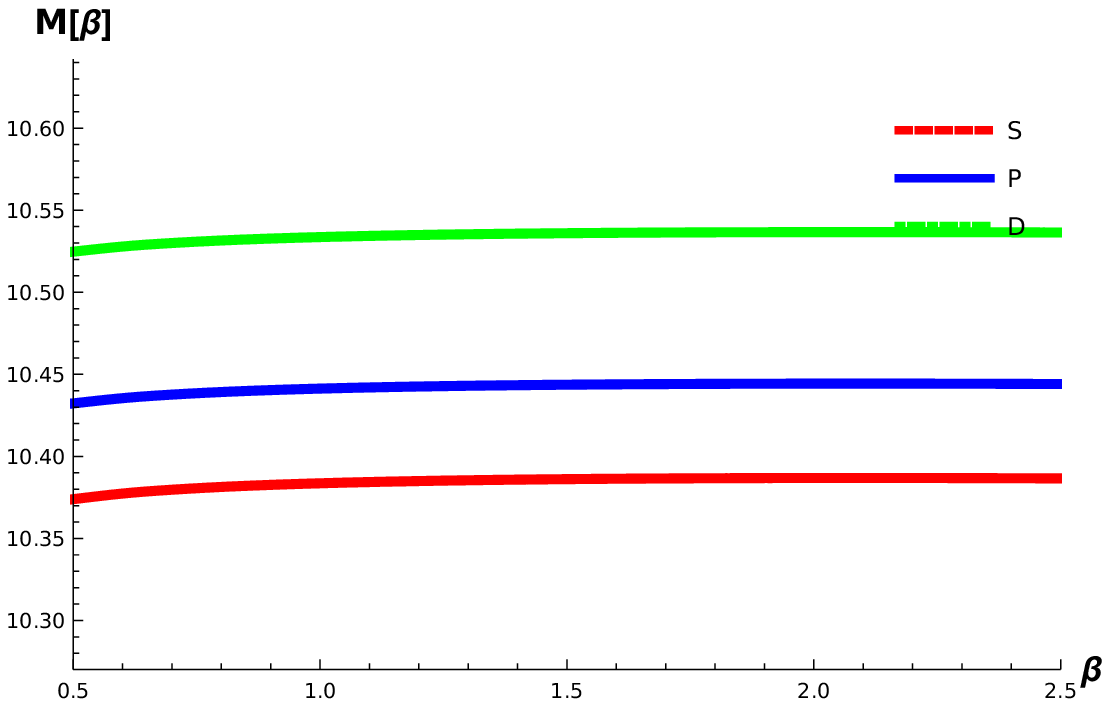}}
	\caption{Mass of spectra for the bottomonium for different values of $L$}
	\label{fig5}
\end{figure}

As can be seen from Figure 5, similar to the charmonium, as the temperature ($T$) is getting smaller ($\beta$ increases) the mass values increase for all values of $n$ for the bottomonium but much higher values.

\begin{figure}[!h]
	\centering 
	\subfigure[Partial function]{\includegraphics[width=5.3cm]{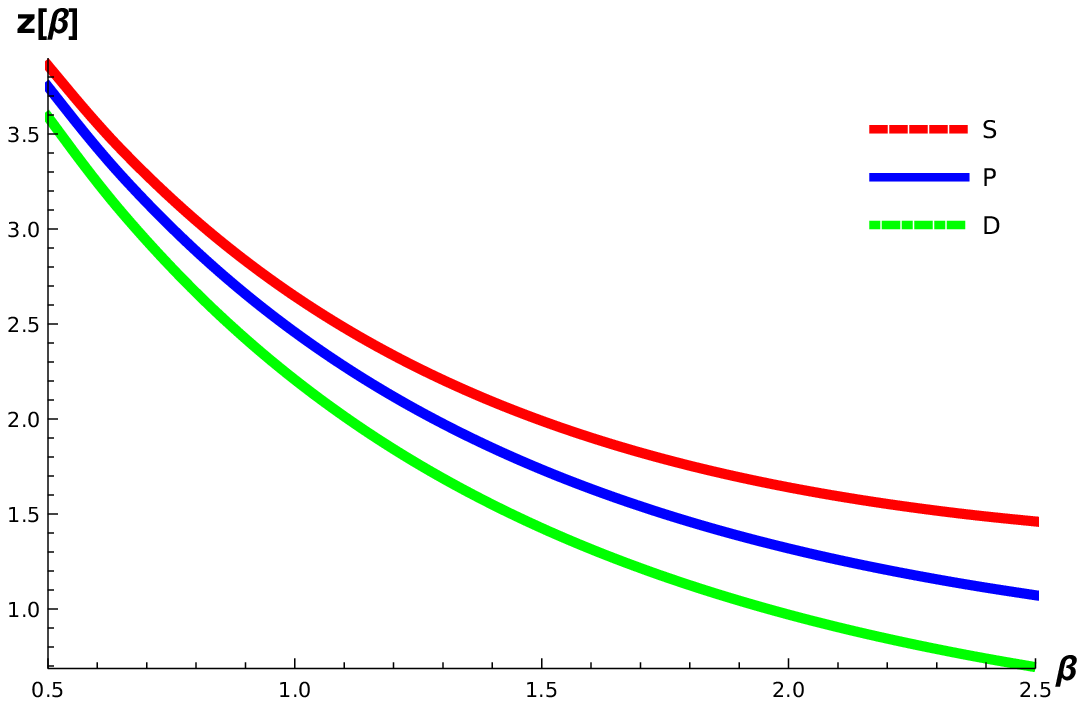}}
	\subfigure[Free Energy]{\includegraphics[width=5.3cm]{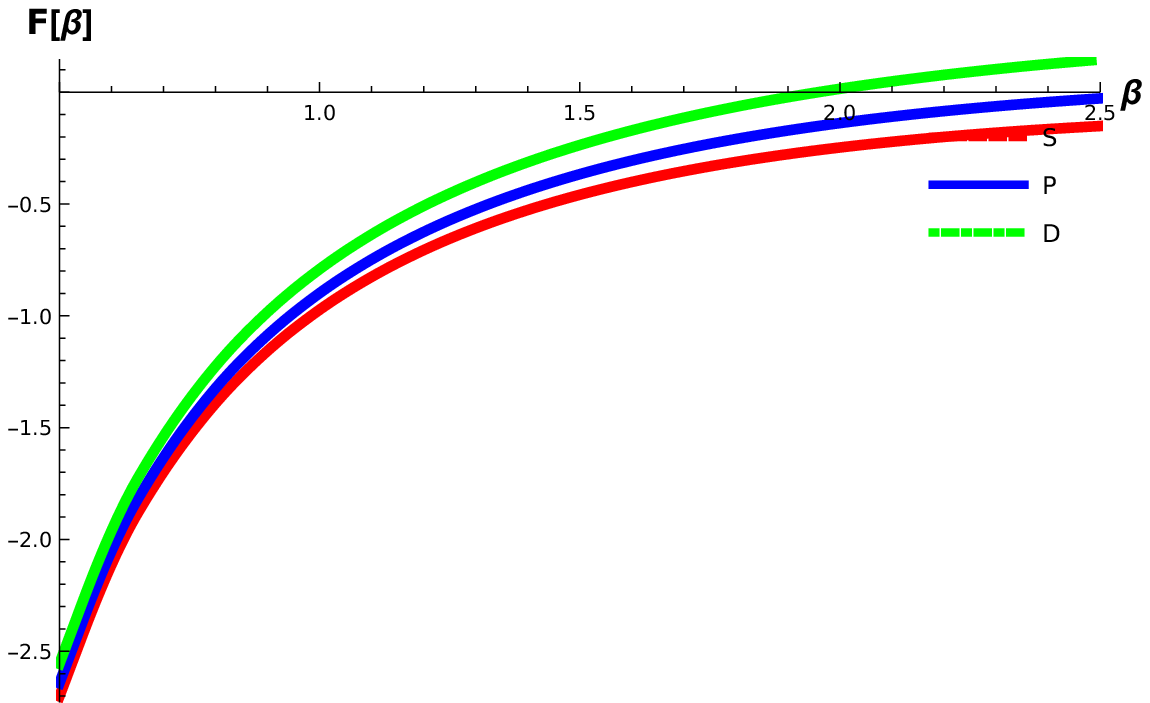}}
	\subfigure[Mean Energy]{\includegraphics[width=5.3cm]{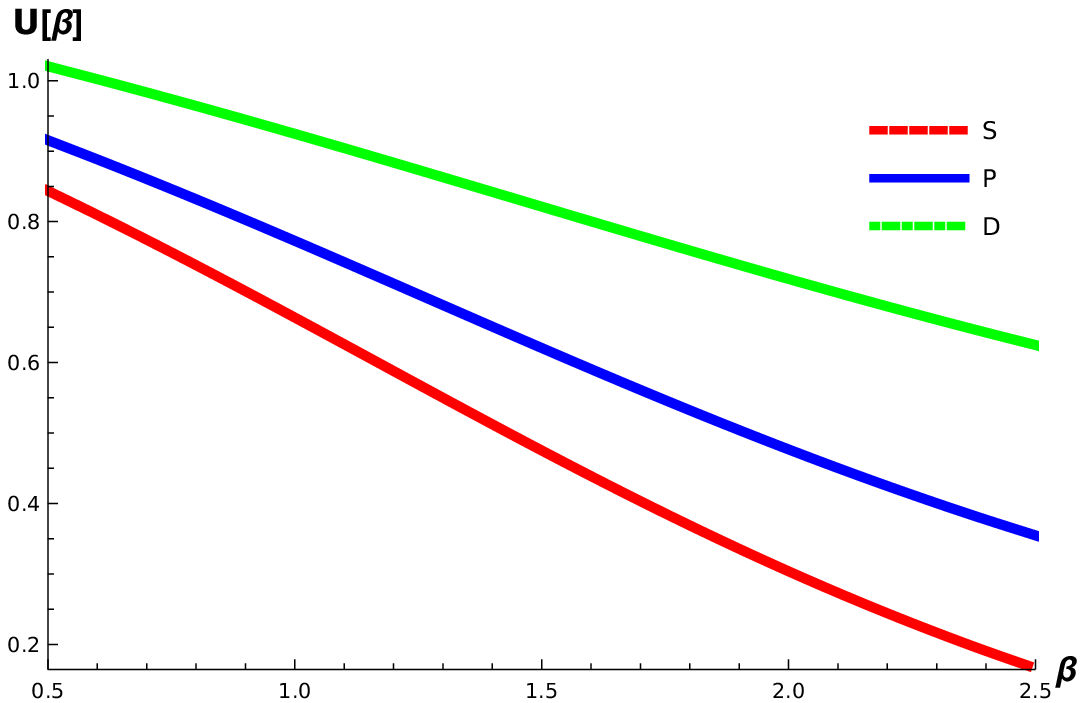}}\\
	\subfigure[Specific Heat]{\includegraphics[width=5.3cm]{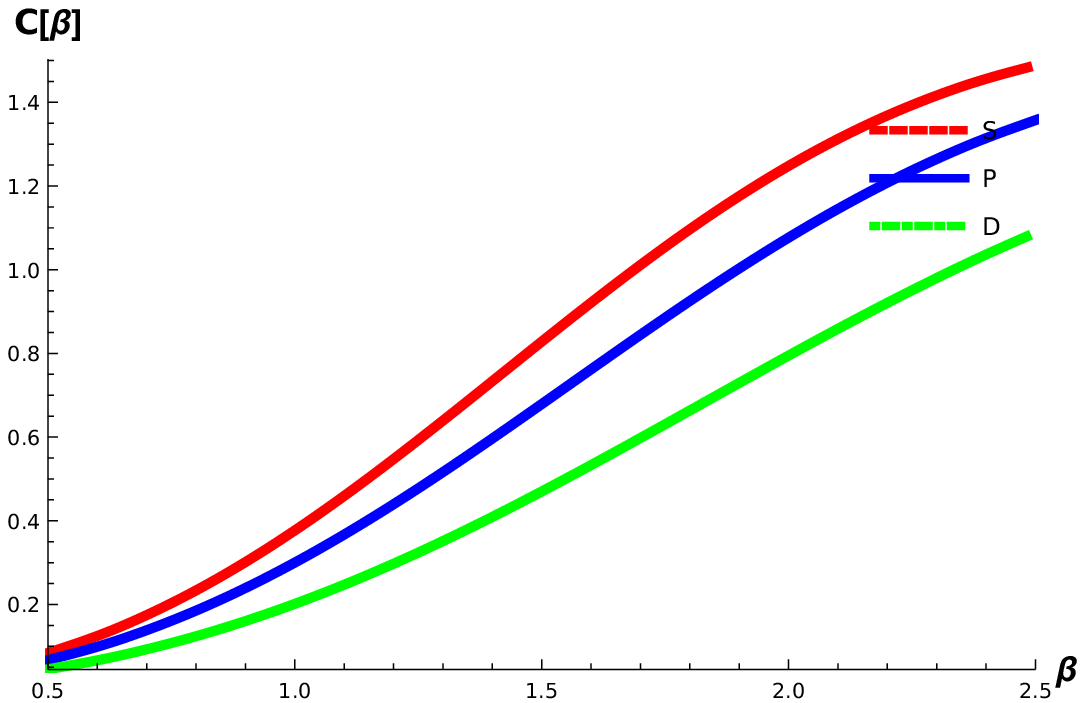}}
	\subfigure[Entropy]{\includegraphics[width=5.3cm]{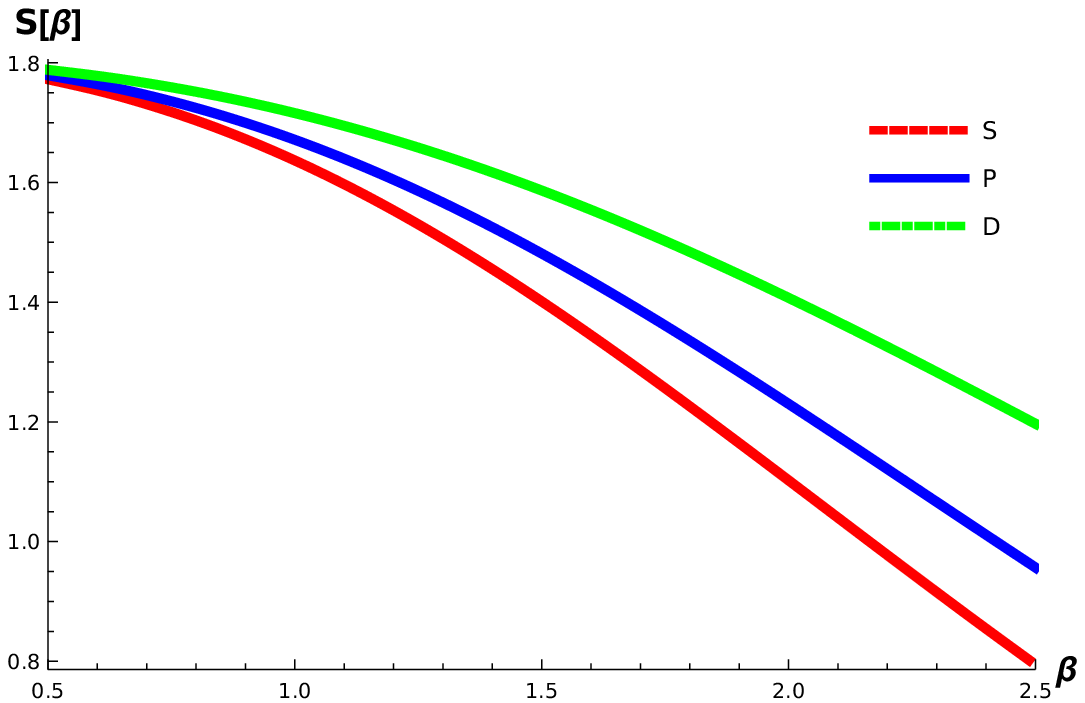}}
	\subfigure[Magnetic Susceptibility]{\includegraphics[width=5.3cm]{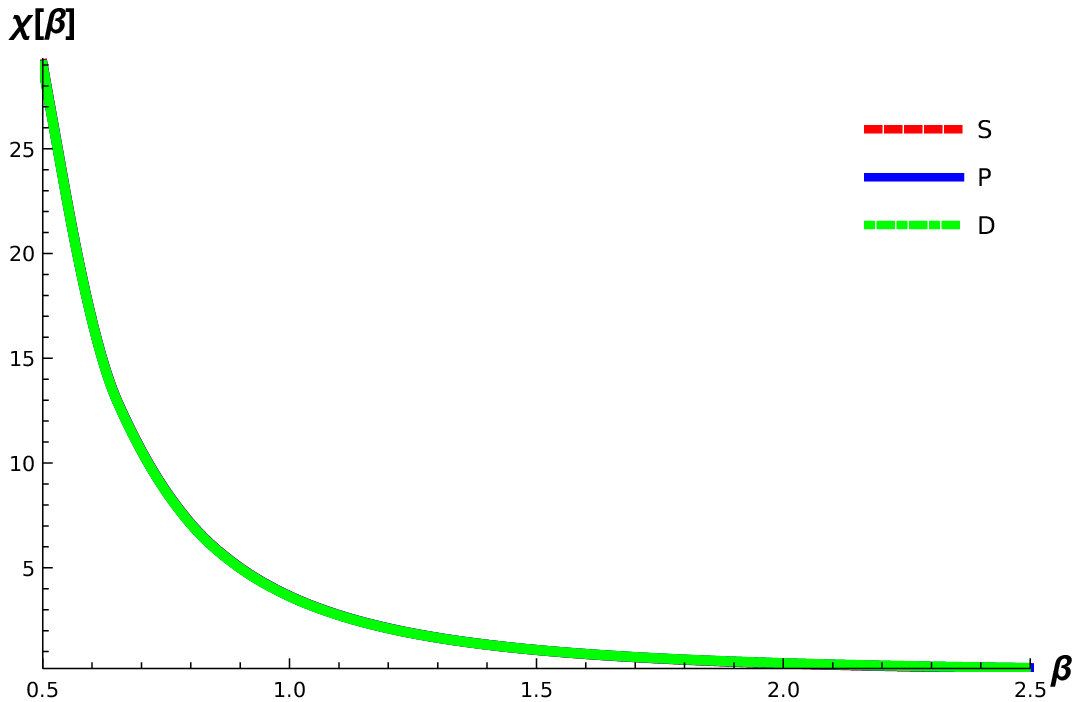}}\\
	\caption{ Thermodynamic quantities for the bottomonium for different values of $L$}
	\label{fig6}
\end{figure}

We displayed the change of the thermodynamic quantities with increasing values of the $\beta$ in Figure 6. Again it is seen that all of them are getting their steady values at a sufficiently large $\beta$ value and they also decrease except the free energy function and specific heat. For the high values of $\beta$, the entropy is getting very small \cite{47}. There is a very little probability of transition to a higher energy level due to the occupation of the lowest level. Also with decreasing values of $\beta$, the entropy is very large. This behavior enforces the probability of the transition to go up. When all levels are occupied at small $\beta$ there is a little change in the entropy for small changes in $\beta$ and a lower heat capacity.   Magnetic susceptibility efficiency drops off rapidly with  $\beta$. As a result, for small $\beta$ values thermal effects are dominant and there is no response for the quantum phenomena in them.

\section{Conclusion}

In this article, the temperature dependent energy, mass and thermodynamic quantities (free energy $F(\beta)$, mean energy $U(\beta)$, specific heat $C(\beta)$, entropy $\mathcal{S}(\beta)$ and magnetic susceptibility $\chi(\beta)$ in terms of canonical partition function $Z(\beta)$) for quarkonia is calculated using the temperature dependent energy eigenvalue expansion. We calculated these quantities using the fixed values of bare quark masses and parameter sets and shown that these results agree with the literature. The effect of the temperature dependent energy eigenvalue expansion can be observed from the small differences in the numerical values compared to the experimental ones that use fixed temperature.The smallness of the difference means that it is  worth introducing the temperature-dependence in the potential. Interesting results are obtained in the case of bottonium and charmonium in the sense of displaying the bound-state mass seems to be less sensitive to temperature changes. Also, the dependence of the thermodynamic quantities on $L$ is seen to be stronger on bottomonium than on charmonium. The results of the temperature dependent energy eigenvalue expansion can be improved by using similar parameter sets ($A, B, C, D$ and $\delta$) used for the experimental setup. Temperature has a positive effect on all of the thermal properties except the free energy. As a further study, it may be very interesting to extend this work to include spin-spin interactions, spin-orbital interaction or complex potential case. Also as another extension, one can consider the Shanon and Renyi entropies for quarkonia. These new results may be used in numerous practical purposes such as measuring the squeezing of quantum fluctuation and reconstruction of the charge and momentum, which are closely related to fundamental and experimentally measurable quantities.

\section*{Data Availability}

No data were used to support this study.

\section*{Conflicts of Interest}

The author declares no conflicts of interest.

\section*{Founding Statement}

The author declare that this research did not receive any funding from any sources.

\end{document}